\documentclass[conference]{IEEEtran}
\IEEEoverridecommandlockouts

\usepackage{cite}
\usepackage{orcidlink}
\usepackage{amsmath,amssymb,amsfonts}
\usepackage{algorithmic}
\usepackage{graphicx}
\usepackage{textcomp} 
\usepackage{xcolor}
\usepackage{algorithm}
\usepackage{pgfplots}
\pgfplotsset{compat=1.18}
\usepackage{tikz}
\usetikzlibrary{positioning}
\usepgfplotslibrary{groupplots}  
\usepackage{comment}

\definecolor{RFFGreen}{RGB}{0,127,0}
\definecolor{StandardOrange}{RGB}{230,102,0}

\def\BibTeX{{\rm B\kern-.05em{\sc i\kern-.025em b}\kern-.08em
    T\kern-.1667em\lower.7ex\hbox{E}\kern-.125emX}}
\begin{document}

\title{\LARGE \bf Tube-Based Model Predictive Control with Random Fourier Features for Nonlinear Systems}

 \author{
 	\parbox{\textwidth}{%
 		\centering
 		\'Akos M. Bokor\orcidlink{0009-0005-4845-9923}, Tam\'as D\'ozsa\orcidlink{0000-0003-0919-4385}, Felix Biert\"umpfel\orcidlink{0000-0002-0016-9637}, \'Ad\'am Szab\'o\orcidlink{0000-0003-1633-5588}
 	}%
    \thanks{The research was supported by the European Union within the framework of the National Laboratory for Autonomous Systems. (RRF-2.3.1-21-2022-00002). The project has received funding from the Swiss Government Excellence Scholarship No. 2025.0057.
    }
 	\thanks{\'A. M. Bokor is with the Systems and Control Laboratory, HUN-REN Institute for Computer Science and Control (SZTAKI), Kende utca 13-17., H-1111~Budapest, Hungary
 		{\tt\small bokor.akos.mark@sztaki.hun-ren.hu}}%
 	\thanks{T. D\'ozsa is with the Systems and Control Laboratory, HUN-REN Institute for Computer Science and Control (SZTAKI), Kende utca 13-17., H-1111~ Budapest, Hungary, and also with Unidistance Suisse, Faculty of Mathematics and Computer Science, Applied mathematics research group, Brig, Switzerland
 		{\tt\small dozsa.tamas@sztaki.hun-ren.hu}}%
    \thanks{F. Biert\"uempfel is with the  Department of Electrical Engineering and Computer Science at the University of Michigan, Ann Arbor, USA. 
 		{\tt\small felixb@umich.edu}}%
    \thanks{\'A. Szab\'o is with the Department of Control for Transportation and Vehicle Systems, Faculty of Transportation Engineering and Vehicle Engineering, Budapest University of Technology and Economics, Műegyetem rkp. 3., H-1111~Budapest, Hungary, and also with the Systems and Control Laboratory, HUN-REN Institute for Computer Science and Control (SZTAKI), Kende utca 13-17., H-1111~Budapest, Hungary
 		{\tt\small szabo.adam@kjk.bme.hu}}%
 }

\maketitle

\begin{abstract}
This paper presents a computationally efficient approach for robust Model Predictive Control of nonlinear systems by combining Random Fourier Features with tube-based MPC. Tube-based Model Predictive Control provides robust constraint satisfaction under bounded model uncertainties arising from approximation errors and external disturbances. The Random Fourier Features method approximates nonlinear system dynamics by solving a numerically tractable least-squares problem, thereby reducing the approximation error. We develop the integration of RFF-based residual learning with tube MPC and demonstrate its application to an autonomous vehicle path-tracking problem using a nonlinear bicycle model. Compared to the linear baseline, the proposed method reduces the tube size by approximately 50\%, leading to less conservative behavior and resulting in around 70\% smaller errors in the test scenario. Furthermore, the proposed method achieves real-time performance while maintaining provable robustness guarantees.
\end{abstract}


\section{Introduction}
Control of nonlinear dynamical systems remains a central challenge in autonomous vehicles, spacecraft operations, and industrial automation. The growing complexity of modern control applications creates increasing demands for approaches that are both safe and computationally efficient.

\emph{Model Predictive Control} (MPC) has emerged as a cornerstone of modern control systems through its capability to handle constraints and optimize performance over a prediction horizon. However, applying MPC to nonlinear systems presents two key challenges: 1) accurately capturing nonlinear dynamics while maintaining computational tractability; 2) ensuring robust constraint satisfaction in the presence of model uncertainties and disturbances.

Learning-based MPC approaches address the modeling challenge by learning system dynamics from data, with \emph{Gaussian Process} (GP) methods emerging as a leading framework. Besides capturing nonlinearities, \emph{Gaussian Process learning-based Model Predictive Control} (GP-MPC) also provides rigorous uncertainty quantification through probabilistic modeling, enabling robust constraint satisfaction via probabilistic constraints. This approach has been successfully demonstrated in safety-critical applications such as autonomous racing \cite{kabzan2019learning}. The standard GP-MPC formulation requires $\mathcal{O}(M^3)$ operations, where $M$ is the number of training data points. The computational bottleneck for kernel methods (including GP-MPC) is the necessity to evaluate the kernel map at every training point. While recent advances address this scaling through dual GP architectures \cite{liu2025dual} or reformulations as quadratic programs \cite{polcz2023efficient}, the cubic complexity remains a challenge for large-scale real-time applications.
There are also other kernel-based approaches: deterministic kernel ridge regression provides finite-sample error bounds without probabilistic assumptions \cite{maddalena2021kernel}, though still requiring kernel evaluations at all training points.

An alternative robust control approach is tube-based MPC, which separates the control problem into two subtasks: steering a nominal trajectory and guaranteeing that the true system trajectory remains within a bounded tube around the nominal one. The tube represents a set around the nominal path, within which the actual state is guaranteed to remain in the presence of uncertainties. For nonlinear systems, tube MPC theory provides formal robustness guarantees through constraint tightening based on uncertainty bounds \cite{schwenkel2022model,yin2020reachability, bokor2025robust}. Learning-based tube MPC combines these guarantees with data-driven models, either through GP-based uncertainty propagation \cite{kohler2021robust,kohler2023robust} or neural network learning \cite{fan2020deep,gao2024learning}. Each approach presents different trade-offs between computational scalability and deterministic uncertainty bounds.

The present paper combines \emph{Random Fourier Features} (RFF) with tube-based MPC to achieve both computational efficiency and formal robustness guarantees. RFF is a kernel-based approximation scheme where the key idea is to approximate the kernel of a large, typically infinite-dimensional reproducing kernel Hilbert space through projection onto a finite-dimensional subspace (also referred to as lifted feature space henceforth)
\cite{rahimi2007}, \cite{liu2021random}. Thus, the obtained approximation can be interpreted as a solution to a least squares problem in the lifted feature space.
The computational advantage is substantial: evaluating RFF features requires $\mathcal{O}(D(n+m))$ operations, where $D$ is the number of features, $n$ is the state dimension, and $m$ is the input dimension. Critically, this complexity is independent of the training set size $M$, avoiding the cubic scaling that limits GP methods.

Learning residual dynamics, corrections to a physics-based linear baseline, rather than modeling the full nonlinear system, has been a popular approach. This residual learning paradigm has proven effective in applications ranging from autonomous vehicles \cite{kabzan2019learning} to real-time disturbance rejection \cite{zhang2025mpc}. The residual $r(x,u) = f(x,u) - (Ax + Bu)$ captures nonlinear effects and linearization errors that the baseline misses. Since the linear model captures dominant dynamics, the residual can often be approximated well using low complexity learning schemes. The explicit $(A,B)$ structure of the combined model integrates naturally with tube MPC theory, while the learned residual dramatically reduces the uncertainty bound $d_{\max}$ compared to using only the linear baseline. This reduction directly decreases conservatism in constraint tightening.

Existing RFF-based control work focuses on nominal MPC performance with online adaptation \cite{zhou2024simultaneous} or tailored feature designs for specific system classes \cite{kazemian2024random}, but does not address robust constraint satisfaction through tubes. Conversely, learning-based tube MPC provides robustness guarantees but relies on computationally expensive GP evaluations or neural networks lacking formal uncertainty quantification. Our approach bridges this gap by integrating RFF efficiency with tube MPC robustness.

Our contributions are threefold:
\begin{enumerate}
    \item  We develop an RFF-based tube MPC solution with deterministic uncertainty quantification, deriving explicit formulas for constraint tightening based on RFF approximation error bounds that enable tube sizing without probabilistic assumptions. 
    \item We maintain exact constraint satisfaction through optimization in physical space with computational complexity $\mathcal{O}(D(n+m))$ per MPC iteration, independent of training set size. \item We validate the proposed approach on an autonomous vehicle path-tracking problem using a nonlinear bicycle model, demonstrating that RFF-based residual learning achieves order-of-magnitude reductions in uncertainty bounds compared to linear-only tube MPC, resulting in significant improvements in tracking performance while maintaining real-time feasibility.

\end{enumerate}

The paper is organized as follows: In Section~\ref{Section 2} the basics of RFF-based residual dynamics learning is described. The MPC tube framework and the proposed approach are detailed in Section~\ref{Section 3}, and Section~\ref{Section 4} shows the effectiveness of the developed method for path tracking using a nonlinear bicycle model. At last, the overall conclusions are presented in Section~\ref{Section 5}. 

\section{RFF-Based Residual Dynamics Learning}\label{Section 2}

We develop an RFF-based approach to learn nonlinear residual dynamics for integration with tube MPC. The method combines a physics-based linear baseline with data-driven corrections. All learning procedures—data generation, model training, and uncertainty quantification—are performed offline prior to deployment. During online operation, the MPC only evaluates the learned model with fixed parameters.

\subsection{Problem Formulation}

Consider the discrete-time nonlinear system:
\begin{equation}
    x_{t+1} = f(x_t, u_t) + w_t,
\end{equation}
where $x_t \in \mathbb{R}^n$ is the state, $u_t \in \mathbb{R}^m$ is the control input, and $w_t \in \mathbb{R^n}, \ (n,m,t \in \mathbb{N})$ represents process noise or unmodeled dynamics.
From the linearization of a physics-based first-principle model, we obtain the linear baseline dynamics:
\begin{equation}
    x_{t+1} = Ax_t + Bu_t,
\end{equation}
where $(A, B) \in \mathbb{R}^{n \times n} \times \mathbb{R}^{n \times m}$.
The residual dynamics, capturing nonlinear effects and linearization errors, are defined as:
\begin{equation}
    r(x, u) := f(x, u) - (Ax + Bu).
    \label{eq:residual equation label}
\end{equation}
Since the linear model captures the dominant dynamics, this residual $r(x,u)$ is typically small over the operational domain, making it amenable to accurate data-driven learning with modest model complexity.

\subsection{Random Fourier Features}

Random Fourier Features provide an explicit finite-dimensional approximation of residual system dynamics (see Eq.~\eqref{eq:residual equation label}) suitable for real-time control. For shift-invariant kernels such as the Radial Basis Function (RBF) kernel:
\begin{equation}
    \zeta(z, z') = \exp\left(-\frac{\|z - z'\|^2}{2\sigma^2}\right),
    \label{Eq 4}
\end{equation}
where $\|\cdot\|_2$ denotes the Euclidean norm, and $\sigma > 0$ is the kernel length scale parameter,
Bochner's theorem \cite{rahimi2007random} enables the construction of an explicit map $\phi: \mathbb{R}^{n+m} \to \mathbb{R}^D$ such that:
\begin{equation}
    \zeta(z, z') \approx \phi(z)^\top \phi(z'),
\end{equation}
where the RFF feature map is:
\begin{equation}
    \phi(z) = \sqrt{\frac{2}{D}} \begin{bmatrix} 
        \cos(\omega_1^\top z + b_1) \\ 
        \vdots \\ 
        \cos(\omega_D^\top z + b_D) 
    \end{bmatrix} \in \mathbb{R}^D.
    \label{eq:rff_features}
\end{equation}
The frequency vectors $\omega_i \in \mathbb{R}^{n+m}$ and phase shifts $b_i \in \mathbb{R}$ are sampled according to:
\begin{equation}
    \omega_i \sim \mathcal{N}(0, \sigma^{-2}I_{n+m}), \quad b_i \sim \text{Uniform}[0, 2\pi], \quad i=1,\ldots,D.
\end{equation}

As a consequence of Bochner's theorem this choice guarantees that $\phi(z)^\top \phi(z')$ from Eq.~\eqref{Eq 4} is an unbiased estimator of the RBF kernel.
The key advantage is that $\phi(z)$ can be computed in $\mathcal{O}(D(n+m))$ time, independent of training set size $M$. For typical choices $D \ll M$, this yields substantial computational savings while maintaining approximation quality.

\subsection{Model Training}

Our goal is to learn the residual mapping $r: \mathbb{R}^{n+m} \to \mathbb{R}^n$ that captures the nonlinear effects and linearization errors of the baseline model
$(A,B)$. We employ Random Fourier Features to approximate this residual function, enabling efficient online MPC evaluation while maintaining approximation accuracy.

The RFF approximation represents the residual as:
\begin{equation}
    r(x, u) \approx W^\top \phi(x, u),
\end{equation}
where $W \in \mathbb{R}^{D \times n}$ is a weight matrix to be learned from data, and $\phi(\cdot)$ is the fixed RFF feature map defined in~(\ref{eq:rff_features}). The training data $\{(z_i, r_i)\}_{i=1}^M$ is generated by simulating the true nonlinear dynamics over the expected operational domain defined by state bounds $x \in \mathcal{X}$ and input bounds $u \in \mathcal{U}$, which capture physical constraints and operational requirements. States and inputs are sampled uniformly over this domain, and for each sample, the residual $r_i = f(x_i, u_i) - (Ax_i + Bu_i)$ is computed from the true and linearized dynamics.

Using the training data and RFF parameters $(\omega, b)$, the feature matrix is constructed:
\begin{equation}
    \Phi = [\phi(z_1), \ldots, \phi(z_M)] \in \mathbb{R}^{D \times M},
\end{equation}
and the residual targets are stacked as $R = [r_1, \ldots, r_M]^\top \in \mathbb{R}^{M \times n}$.

The weight matrix $W$ is obtained by solving the ridge regression problem:
\begin{equation}
    W^* = \arg\min_{W} \sum_{i=1}^M \|r_i - W^\top \phi(z_i)\|_2^2 + \lambda \|W\|_F^2,
\end{equation}
which has the closed-form solution:
\begin{equation}
    W^* = (\Phi \Phi^\top + \lambda I_D)^{-1} \Phi R \in \mathbb{R}^{D \times n},
\end{equation}
where $\lambda > 0$ is the regularization parameter.
Combining the linear baseline with the learned residual correction, the complete hybrid dynamics model becomes:

\begin{equation}
    \hat{f}(x, u) = Ax + Bu + W^\top \phi(x, u).
    \label{eq:hybrid_model}
\end{equation}

\subsection{Uncertainty Quantification}

To enable robust constraint satisfaction in tube MPC, we compute a deterministic upper bound on the prediction error:
\begin{equation}
    \|f(x, u) - \hat{f}(x, u)\| \leq d_{\max} \quad \forall (x,u) \in \mathcal{X} \times \mathcal{U}.
    \label{eq:error_bound}
\end{equation}

Since the true function $f$ is unknown analytically, we estimate this bound empirically using a validation set $\{(z_j^{\text{val}}, r_j^{\text{val}})\}_{j=1}^{N_{\text{val}}}$ generated separately from the training data by uniformly sampling over $\mathcal{X} \times \mathcal{U}$. The prediction error for each validation sample is:
\begin{equation}
    e_j = \|r_j^{\text{val}} - W^\top\phi(z_j^{\text{val}})\|_2, \quad j=1,\ldots,N_{\text{val}}.
\end{equation}
The error bound is then computed as:
\begin{equation}
    d_{\max} = \beta \cdot \max_{j=1,\ldots,N_{\text{val}}} e_j,
\end{equation}
where $\beta > 1$ is a safety factor accounting for potential model errors in regions not covered by the validation set.

\section{Tube-Based Robust MPC with RFF Dynamics}\label{Section 3}

We integrate the RFF-based residual model from Section~\ref{Section 2} into a tube MPC framework following \cite{schwenkel2022model}. The RFF-enhanced dynamics enable tighter uncertainty bounds compared to using only the linear baseline, reducing conservatism while maintaining formal robustness guarantees.

\subsection{Nominal Prediction Model and Error Dynamics}

The $\xi_{t|k}\in\mathbb{R}^{n}$ trajectory of the nominal, uncertainty-free model is governed by:
\begin{subequations}
\label{eq:nominal_dynamics}
\begin{align}
\xi_{t+1|k} &= A \xi_{t|k} + B v_{t|k} + W^\top \phi(\xi_{t|k}, v_{t|k}), \\
r_{t|k} &= C \xi_{t|k} + D v_{t|k},
\end{align}
\end{subequations}
where $\xi_{t|k}$ is the nominal state predicted at $t$ steps into the future from starting time $k\in\mathbb{Z}$ if the nominal inputs $v_{t|k}$ act on the system. At time $k$, the optimization problem includes inputs $v_{t|k}$ for $t \in \{\tau_0, \tau_1, \ldots, \tau_{N-1}\}$.

Here, $A$ and $B$ represent the linear baseline dynamics from Section~\ref{Section 2}, while $W^\top \phi(\cdot)$ captures the learned nonlinear residual. The matrices $C \in \mathbb{R}^{n_\mathrm{r} \times n}$ and $D \in \mathbb{R}^{n_\mathrm{r} \times m}$ define the nominal output $r_{t|k}$. For state feedback, we define $C = I_n$ and $D = 0$, such that $r_{t|k}$ represents the nominal state trajectory used in tube size propagation.

The actual input applied to the system is the nominal input determined by the MPC and a correction term calculated through a stabilizing feedback gain $K\in\mathbb{R}^{m\times n}$:
\begin{equation}
    u_{t|k} = v_{t|k} + K e_{t|k},
    \label{eq:control_law}
\end{equation}
with, $e_{t|k} = x_{t|k} - \xi_{t|k}$ being the error between the true and nominal states. The gain $K$ is designed to ensure that $(A + BK)$ is Schur stable, typically using a \emph{Linear Quadratic Regulator} (LQR):
\begin{equation}
    K = -\operatorname{dlqr}(A, B, Q_K, R_K),
\end{equation}
for appropriately chosen weighting matrices $Q_K \succeq 0$ and $R_K \succ 0$.

By substituting the control law \eqref{eq:control_law} into the true system dynamics $x_{t+1} = f(x_t, u_t)$ and using the RFF approximation $f(x,u) \approx \hat{f}(x,u)$ from \eqref{eq:hybrid_model}, we obtain the error dynamics:
\begin{subequations}
\label{eq:error_dynamics}
\begin{align}
    e_{t+1|k} &= A_\mathrm{e} e_{t|k} + d_{t|k}, \\
    z_{t|k} &= C_\mathrm{e} e_{t|k} + r_{t|k},
\end{align}
\end{subequations}
where $A_\mathrm{e}:= A + BK$, $C_\mathrm{e}:= C + DK$, and $d_{t|k}$ takes into account the combined effect of RFF approximation error and higher-order linearization terms. For small tracking errors where the RFF model is accurate, this disturbance term satisfies:
\begin{equation}
    \|d_{t|k}\| \leq d_{\max},
    \label{eq:disturbance_bound}
\end{equation}
where $d_{\max}$ is the error bound computed in Section~\ref{Section 2}.

\subsection{Robust Constraint Tightening}

The polytopic constraints are defined with respect to the state $x_k$ and input $u_k$ in the following form:
\begin{equation}
H \begin{bmatrix} x_k \\ u_k \end{bmatrix} \leq h,
\label{eq:polytopic_constraints}
\end{equation}
where $H \in \mathbb{R}^{n_\mathrm{h} \times (n+m)}$ and $h \in \mathbb{R}^{n_\mathrm{h}}$.

However, the MPC optimization uses nominal variables $\xi_{t|k}$ and $v_{t|k}$, which are related to the true variables by:
\begin{equation}
    x_{t|k} = \xi_{t|k} + e_{t|k}, \quad u_{t|k} = v_{t|k} + K e_{t|k}.
\end{equation}

To guarantee robust constraint satisfaction, we apply constraint tightening based on a quadratic bound on the tracking error:
\begin{equation} \label{eq:tube}
    \|e_{t|k}\|_P^2 = e_{t|k}^\top P e_{t|k} \leq s_{t|k},
\end{equation}
where $s_{t|k}$ is a scalar sequence describing the tube size at each time $t$, and $P \succ 0$ is an arbitrary positive definite matrix.

Using the bound on $e_{t|k}$, the $i^\mathrm{th}$ constraint takes the following tightened form:
\begin{equation}
    H_i \begin{bmatrix} \xi_{t|k} \\ v_{t|k} \end{bmatrix} \leq h_i - g_i \sqrt{s_{t|k}},
    \label{eq:tightened_constraints}
\end{equation}
where $g_{i}$ is the tightening factor of the $i^\mathrm{th}$ constraint, which is defined by:
\begin{equation} \label{eq:tight}
    g_i := \left\| H_i \begin{bmatrix} I \\ K \end{bmatrix} P^{-1/2} \right\|_2.
\end{equation}

Satisfaction of the original constraints~(\ref{eq:polytopic_constraints}) is guaranteed for all trajectories within the tube defined by $P$ and $s_{t|k}$ if the nominal trajectory satisfies the tightened constraints~(\ref{eq:tightened_constraints}).


\subsection{Tube Size Evolution}

Comptuting the evolution of the tube size $s_{t|k}$ is a key component for tube-MPC. For the error dynamics \eqref{eq:error_dynamics} with bounded disturbance $\|d_{t|k}\| \leq d_{\max}$, we have:
\begin{equation}
    e_{t+1|k}^\top P e_{t+1|k} \leq \rho^2 \, e_{t|k}^\top P e_{t|k} + \Xi d_{\max}^2,
\end{equation}
where $\Xi > 0$ is a disturbance scaling factor and $\rho \in (0,1)$ characterizes the exponential stability of the error dynamics .
This yields the tube size recursion:
\begin{equation}
s_{t+1|k} = \rho^2 s_{t|k} + \Xi d_{\max}^2.
\label{eq:s_update}
\end{equation}

The tube size is initialized at the start of each MPC update as follows:
\begin{equation}
s_{0|k} = s_{1|k-1} + e_{0|k}^\top P e_{0|k} - e_{1|k-1}^\top P e_{1|k-1},
\label{eq:s_init} 
\end{equation}
where $e_{0|k} = x_{0|k} - \xi_{0|k}$ with $x_{0|k}$ being the actual measured state and $\xi_{0|k}$ being the nominal trajectory at the current timestep, and $e_{1|k-1}$ is the deviation from the real trajectory for time $k$, which was predicted at the previous step.

\subsection{Terminal Set and Recursive Feasibility}

To guarantee recursive feasibility and closed-loop stability, the nominal trajectory and tube size must reach and remain within a terminal set where a terminal controller can maintain constraint satisfaction indefinitely.

Hence, a terminal controller $K_\Omega$ must be designed such that $(A + BK_\Omega)$ is Schur stable. The terminal cost $S \succ 0$ is obtained from the discrete-time Lyapunov equation:
\begin{equation}
(A+B K_\Omega)^\top S (A+B K_\Omega) - S = -(Q + K_\Omega^\top R K_\Omega),
\end{equation}
which ensures the required decrease in terminal cost.

The terminal set is defined as:
\begin{equation}
\Omega = \{ (\xi,s)\in\mathbb{R}^{n} \times \mathbb{R} \mid \|\xi\|_S^2 \leq \gamma_1, \ 0 \leq s \leq \gamma_2 \},
\label{eq:terminal_set}
\end{equation}
The parameters $\gamma_1, \gamma_2 > 0$ are designed to guarantee two properties within the terminal set $\Omega$: robust positive invariance and constraint satisfaction. These are are computed by maximizing $\gamma_1$ subject to the existence of $\gamma_2$ satisfying both tube propagation and constraint feasibility within $\Omega$.

\subsection{Online MPC Optimization}

All components of the proposed MPC scheme are now defined, and we can pose the corresponding optimization problem. 
For each $k \geq 0$, the state $x_k$ is obtained through measurement, after which the finite-horizon MPC optimization is solved over $\xi_{t|k}$, $v_{t|k}$, and  $s_{t|k}$ for $t \in \{\tau_0, \tau_1, \ldots, \tau_{N-1}\}$:
\begin{align}
\min_{\xi, v, s} \quad & \sum_{t=0}^{N-1} \left(\|\xi_{t|k}\|_Q^2 + \|v_{t|k}\|_R^2\right) + \|\xi_{N|k}\|_S^2 \label{eq:main_MPC} \\
\text{s.t.} \quad & \text{Nominal dynamics }~(\ref{eq:nominal_dynamics}), \notag \\
& \text{Tube size evolution }~\ref{eq:s_update}),~(\ref{eq:s_init}),  \notag \\
& \text{Tightened constraints }~(\ref{eq:tightened_constraints}), \notag \\
& \text{Terminal set }~(\ref{eq:terminal_set}). \notag
\end{align}

In this formulation, the stage cost consists of penalties on the nominal state $Q \succeq 0$ and on the control effort $R \succ 0$. Instead of fixing it to the measured state $x_k$, the initial nominal state $\xi_{0|k}$ is included in the optimization problem as a decision variable, which leads to reduced conservatism.

After solving \eqref{eq:main_MPC}, according to the receding horizon principle, only the first step of the solution is applied to the system:
\begin{equation}
    u_k = v_{0|k}^\star + K (x_k - \xi_{0|k}^\star), \label{eq:comp_input}
\end{equation}
where the superscript $^\star$ denotes the decision variables found by solving \eqref{eq:main_MPC} at time $k$.

\subsection{Reduced Conservatism via RFF Learning}

The advantage of RFF-based residual learning is quantified through the uncertainty reduction:
\begin{equation}
\frac{d_{\max}^{\mathrm{rff}}}{d_{\max}^{\mathrm{lin}}} \ll 1,
\end{equation}
where $d_{\max}^{\mathrm{lin}}$ corresponds to using only the linear baseline and $d_{\max}^{\mathrm{rff}}$ includes the learned residual.

Since constraint tightening scales as $g_i\sqrt{s_{\infty}}$ where $s_{\infty} = \Xi d_{\max}^2/(1-\rho^2)$, this reduction directly decreases conservatism:
\begin{equation}
\frac{\text{Tightening}^{\mathrm{rff}}}{\text{Tightening}^{\mathrm{lin}}} = \frac{d_{\max}^{\mathrm{rff}}}{d_{\max}^{\mathrm{lin}}}.
\end{equation}
The approach thus maintains formal robustness guarantees while enabling operation closer to constraint boundaries.

\section{Application to Path Tracking Using a Nonlinear Bicycle Model}\label{Section 4}

We demonstrate the proposed RFF-based tube MPC on a nonlinear kinematic bicycle model for path-tracking with aggressive maneuvering requirements.

The bicycle model describes lateral deviation $e_y$ and heading error $e_\psi$ from a reference path:
\begin{subequations}
\begin{align}
    e_y(k+1) &= e_y(k) + v\sin(e_\psi(k))\Delta t, \\
    e_\psi(k+1) &= e_\psi(k) + \frac{v}{L}\tan(\delta(k))\Delta t - v\kappa(k) \Delta t,
\end{align}
\end{subequations}
where $v$ is the velocity, $L$ is the wheelbase, $\delta$ is the steering angle, and $\kappa$ is the path curvature. Linearization around $e_\psi \approx 0$, $\delta \approx 0$ yields:
\begin{multline}
    \begin{bmatrix} e_y(k+1) \\ e_\psi(k+1) \end{bmatrix} = \underbrace{\begin{bmatrix} 1 & v\Delta t \\ 0 & 1 \end{bmatrix}}_{A} \begin{bmatrix} e_y(k) \\ e_\psi(k) \end{bmatrix} \\ + \underbrace{\begin{bmatrix} 0 \\ \frac{v}{L}\Delta t \end{bmatrix}}_{B} \delta(k) + \begin{bmatrix} 0 \\ -v\kappa(k)\Delta t \end{bmatrix},
\end{multline}
with the residual term capturing the trigonometric nonlinearities:
\begin{equation}
    r(e_\psi(k), \delta(k)) = \begin{bmatrix} v(\sin(e_\psi(k)) - e_\psi(k))\Delta t \\ \frac{v}{L}(\tan(\delta(k)) - \delta(k))\Delta t \end{bmatrix}.
\end{equation}
The reference trajectory is a slalom maneuver with growing amplitude, resulting in increasingly sharper turns, designed  to induce growing nonlinearities and emphasize the impact of linearization errors. Training and validation data were sampled uniformly over the operational domain $e_y \in [-6, 6]$ [m], $e_\psi \in [-0.8, 0.8]$ [rad], $\delta \in [-0.6, 0.6]$ [rad], and $\kappa \in [-0.4, 0.4]$ [rad/m], defined based on physical vehicle limitations and an additional safety margin. The RFF model was trained via ridge regression as described in Section~\ref{Section 2}, yielding an order-of-magnitude reduction in $d_{\max}$ compared to the linear-only baseline.

The tube MPC constraints $|e_y| \leq 1$ [m], $|e_\psi| \leq 0.2$ [rad], and $|\delta| \leq 0.5$ [rad] were chosen similarly to the training domain, representing physical limits on steering angle as well as realistic lateral and heading error bounds (e.g., lane width). Since the MPC constraints lie within the RFF training domain, any feasible MPC trajectory operates in regions where the learned model has been validated, providing additional confidence in the computed error bound $d_{\max}$ during closed-loop operation.

\begin{figure}[htbp]
\centering
\begin{tikzpicture}
\begin{semilogyaxis}[
    width = 0.95\columnwidth,
    height = 0.45\columnwidth,
    xlabel={Time [s]},
    ylabel={Tube Size},
    xmin=0, xmax=12,
    ymin=1e-4, ymax=2e-1,
    xmajorgrids=true,
    ymajorgrids=true,
    grid style={densely dotted, black!20},
    axis line style={thick},
    tick style={thick},
legend style={
    at={(0.5,1.05)},              
    anchor=south,
    legend columns=2,              
    draw=none,                     
    fill=none,
    font=\scriptsize,
    column sep=10pt
},
    legend cell align={left},
    tick label style={font=\small},
    label style={font=\normalsize},
    xlabel style={yshift=4pt},      
    ylabel style={yshift=-5pt},     
    enlarge x limits=0.02,         
    enlarge y limits=false,
]

\addplot[
    color={rgb,255:red,0;green,127;blue,0},
    line width=1.8pt,
    solid
] table[x index=0, y index=3, col sep=comma] {bicycle_results.csv};

\addplot[
    color={rgb,255:red,230;green,102;blue,0},
    line width=1.8pt,
    dashed,
    opacity=0.7,
    forget plot                    
] table[x index=4, y index=7, col sep=comma] {bicycle_results.csv};

\end{semilogyaxis}
\end{tikzpicture}
\vspace{-2mm}
\caption{Tube size evolution showing adaptive uncertainty quantification: RFF-Based ({\color[rgb]{0,0.498,0}\rule[0.5ex]{12pt}{2.5pt}}), Traditional ({\color[rgb]{0.9314, 0.58, 0.3}\rule[0.5ex]{3pt}{2.5pt}\hspace{1pt}\rule[0.5ex]{3pt}{2.5pt}\hspace{1pt}\rule[0.5ex]{3pt}{2.5pt}}).}
\label{fig:tube_size}
\end{figure}

The RFF approach achieves a substantial reduction in tube size, as shown in Figure~\ref{fig:tube_size}. On average, the tube size of the RFF-MPC is approximately 50.3\% smaller than the linear baseline. As expected, the RFF-augmented linear model closely represents the original nonlinear dynamics, yielding tighter error bounds $d_{\max}$. This decrease in tube size directly reduces constraint tightening via \eqref{eq:tightened_constraints}, making the RFF-based MPC significantly less conservative than the linear-only approach.

\begin{figure}[htbp]
\centering
\begin{tikzpicture}
\begin{axis}[
    width = 0.95\columnwidth,
    height = 0.45\columnwidth,
    xlabel={Time [s]},
    ylabel={$\delta$ [rad]},
    xmin=0, xmax=12,
    ymin=-0.5, ymax=0.5,
    xmajorgrids=true,
    ymajorgrids=true,
    grid style={densely dotted, black!20},
    axis line style={thick},
    tick style={thick},
    legend style={
        at={(0.5,1.05)},
        anchor=south,
        legend columns=2,
        draw=none,
        fill=none,
        font=\scriptsize,
        column sep=10pt
    },
    legend cell align={left},
    tick label style={font=\small},
    label style={font=\normalsize},
    xlabel style={yshift=4pt},
    ylabel style={yshift=-5pt},
]

\addplot[
    color={rgb,255:red,0;green,127;blue,0},
    line width=1.8pt,
    solid                           
] table[x index=0, y index=1, col sep=comma] {bicycle_control.csv};

\addplot[
    color={rgb,255:red,230;green,102;blue,0},
    line width=1.8pt,
    densely dashed,                 
    forget plot
] table[x index=2, y index=3, col sep=comma] {bicycle_control.csv};

\end{axis}
\end{tikzpicture}
\vspace{-2mm}
\caption{Steering angle command: RFF-Based ({\color[rgb]{0,0.498,0}\rule[0.5ex]{12pt}{2.5pt}}), Traditional ({\color[rgb]{0.9314, 0.58, 0.3}\rule[0.5ex]{3pt}{2.5pt}\hspace{1pt}\rule[0.5ex]{3pt}{2.5pt}\hspace{1pt}\rule[0.5ex]{3pt}{2.5pt}}).}\label{fig:steering}
\end{figure}

The reduced conservatism is reflected in the control signal shown in Figure~\ref{fig:steering}. The RFF-based controller uses steering commands closer to the physical limits during aggressive maneuvers, whereas the linear baseline remains more conservative due to larger uncertainty bounds. The results also show that the RFF approach achieves this without violating constraints, validating that the computed $d_{\max}$ provides reliable error bounds.

This less conservative control directly translates into better tracking performance. Figures~\ref{fig:lateral_error} and~\ref{fig:heading_error} show lateral and heading errors throughout the slalom. The linear baseline shows growing errors as the reference amplitude increases. This is as expected because the nonlinearity becomes more significant, forcing the controller to be more conservative and reducing performance. In contrast, the RFF-based tube MPC maintains tight tracking, with average lateral position and heading errors approximately 74\% and 68\% lower than those of the linear baseline, respectively. The performance gap widens progressively as the slalom amplitude increases, demonstrating that the RFF model effectively captures residual dynamics in operating regions where the linear baseline becomes less accurate.

\begin{figure}[htbp]
\centering
\begin{tikzpicture}
\begin{axis}[
    width = 0.95\columnwidth,
    height = 0.45\columnwidth,
    xlabel={Time [s]},
    ylabel={$e_y$ [m]},
    xmin=0, xmax=12,
    ymin=-0.35, ymax=0.35,
    ytick={-0.3,-0.15,0,0.15,0.3},
    xmajorgrids=true,
    ymajorgrids=true,
    grid style={densely dotted, black!20},
    axis line style={thick},
    tick style={thick},
    legend style={
        at={(0.5,1.05)},              
        anchor=south,
        legend columns=2,              
        draw=none,                     
        fill=none,
        font=\scriptsize,
        column sep=10pt
    },
    legend cell align={left},
    tick label style={font=\small},
    label style={font=\normalsize},
    xlabel style={yshift=4pt},      
    ylabel style={yshift=-5pt},     
    enlarge x limits=0.02,
    scaled ticks=false,
    /pgf/number format/.cd,
        fixed,
        precision=2,
]

\addplot[
    color={rgb,255:red,0;green,127;blue,0},
    line width=1.8pt,
    solid
] table[x index=0, y index=1, col sep=comma] {bicycle_results.csv};

\addplot[
    color={rgb,255:red,230;green,102;blue,0},
    line width=1.8pt,
    dashed,
    dash pattern=on 4pt off 2pt,
    opacity=0.7,
    forget plot                    
] table[x index=4, y index=5, col sep=comma] {bicycle_results.csv};

\end{axis}
\end{tikzpicture}
\vspace{-2mm}
\caption{Lateral position error ($e_y$): RFF-Based ({\color[rgb]{0,0.498,0}\rule[0.5ex]{12pt}{2.5pt}}), Traditional ({\color[rgb]{0.9314, 0.58, 0.3}\rule[0.5ex]{3pt}{2.5pt}\hspace{1pt}\rule[0.5ex]{3pt}{2.5pt}\hspace{1pt}\rule[0.5ex]{3pt}{2.5pt}}).}
\label{fig:lateral_error}

\end{figure}

Importantly, these performance improvements come at modest computational cost. The MPC optimization is solved using CasADi \cite{andersson2019casadi} with the IPOPT solver \cite{wachter2006implementation}. With $D = 300$ RFF features, the average computational time per MPC iteration is 26.3 ms, compared to 14.2 ms for the linear baseline, and is reduced to 21.4 ms with $D = 200$, while maintaining substantial performance gains. Both configurations remain well within real-time feasibility for the 33 ms sampling period, with the $\mathcal{O}(D(n+m))$ overhead per prediction step readily justified by the improvements in tracking accuracy and reduced conservatism. Note that this MATLAB/CasADi implementation 
with IPOPT is intended for comparison purposes, snd was executed on a MacBook Air equipped with an Apple M4 chip and 16 GB unified memory. For real-time applications, specialized embedded solvers such as ACADOS would provide faster computation.

\begin{figure}[htbp]
\centering
\begin{tikzpicture}
\begin{axis}[
    width = 0.95\columnwidth,
    height = 0.45\columnwidth,
    xlabel={Time [s]},
    ylabel={$e_{\psi}$ [rad]},
    xmin=0, xmax=12,
    ymin=-0.17, ymax=0.16,
    ytick={-0.12,-0.06,0,0.06,0.12},
    xmajorgrids=true,
    ymajorgrids=true,
    grid style={densely dotted, black!20},
    axis line style={thick},
    tick style={thick},
    legend style={
        at={(0.5,1.05)},              
        anchor=south,
        legend columns=2,              
        draw=none,                     
        fill=none,
        font=\scriptsize,
        column sep=10pt
    },
    legend cell align={left},
    tick label style={font=\small},
    label style={font=\normalsize},
    xlabel style={yshift=4pt},      
    ylabel style={yshift=-5pt},     
    scaled ticks=false,
    /pgf/number format/.cd,
        fixed,
        precision=2,
]

\addplot[
    color={rgb,255:red,0;green,127;blue,0},
    line width=1.8pt,
    solid
] table[x index=0, y index=2, col sep=comma] {bicycle_results.csv};

\addplot[
    color={rgb,255:red,230;green,102;blue,0},
    line width=1.8pt,
    dashed,
    dash pattern=on 4pt off 2pt,
    opacity=0.7,
    forget plot                    
] table[x index=4, y index=6, col sep=comma] {bicycle_results.csv};

\end{axis}
\end{tikzpicture}
\vspace{-2mm}
\caption{Heading angle error ($e_{\psi}$) demonstrating reduced orientation deviation: RFF-Based ({\color[rgb]{0,0.498,0}\rule[0.5ex]{12pt}{2.5pt}}), Traditional ({\color[rgb]{0.9314, 0.58, 0.3}\rule[0.5ex]{3pt}{2.5pt}\hspace{1pt}\rule[0.5ex]{3pt}{2.5pt}\hspace{1pt}\rule[0.5ex]{3pt}{2.5pt}}).}
\label{fig:heading_error}
\end{figure}

\section{Conclusion} \label{Section 5}

This paper presented a computationally efficient framework for robust tube-based MPC of nonlinear systems by integrating Random Fourier Features with deterministic uncertainty quantification. The RFF approximation enables learning of residual dynamics with $\mathcal{O}(D(n+m))$ complexity per MPC iteration, independent of training set size, while providing explicit error bounds $d_{\max}$ for tube sizing without probabilistic assumptions. The bicycle path-tracking demonstration showed order-of-magnitude reductions in both uncertainty bounds and tube sizes compared to linear-only approaches, translating to substantially improved tracking performance with modest computational overhead.

As the first work combining RFF-based learning with tube MPC guarantees, this represents a proof-of-concept demonstrating the viability of the approach. Future research directions include comprehensive benchmarking against GP-based tube MPC methods and other state-of-the-art learning-based robust controllers to quantify performance trade-offs across different problem classes. Additional directions include further computational optimization through tailored RFF feature designs, online adaptation mechanisms for time-varying dynamics, and extensions to broader classes of nonlinear systems beyond residual learning frameworks.

\bibliographystyle{IEEEtran}
\bibliography{references}

@inproceedings{rahimi2007,
  title={Random features for large-scale kernel machines},
  author={Rahimi, Ali and Recht, Benjamin},
  booktitle={Advances in Neural Information Processing Systems},
  volume={20},
  year={2007}
}

@article{kabzan2019learning,
  title={Learning-based model predictive control for autonomous racing},
  author={Kabzan, Juraj and Hewing, Lukas and Liniger, Alexander and Zeilinger, Melanie N},
  journal={IEEE Robotics and Automation Letters},
  volume={4},
  number={4},
  pages={3363--3370},
  year={2019},
  publisher={IEEE}
}

@inproceedings{maddalena2021kernel,
  title={Learning robust predictive control from noisy input-output data with finite-sample guarantees},
  author={Maddalena, Emilio T and Jones, Colin N and Morari, Manfred},
  booktitle={Proceedings of Machine Learning Research},
  volume={144},
  pages={1--12},
  year={2021},
  organization={PMLR}
}

@article{zhou2024simultaneous,
  title={Simultaneous system identification and model predictive control with no dynamic regret},
  author={Zhou, Yingying and Tzoumas, Vasileios},
  journal={arXiv preprint arXiv:2407.04143},
  year={2024}
}

@inproceedings{kazemian2024random,
  title={Random features approximation for control-affine systems},
  author={Kazemian, Matin and Sattar, Sleiman and Dean, Sarah},
  booktitle={Proceedings of Machine Learning Research},
  volume={242},
  pages={732--744},
  year={2024},
  organization={PMLR}
}

@article{kohler2021robust,
  title={A robust adaptive model predictive control framework for nonlinear uncertain systems},
  author={K{\"o}hler, Johannes and M{\"u}ller, Matthias A and Allg{\"o}wer, Frank},
  journal={International Journal of Robust and Nonlinear Control},
  volume={31},
  number={17},
  pages={8698--8717},
  year={2021},
  publisher={Wiley}
}

@article{kohler2023robust,
  title={Robust adaptive {MPC} using control contraction metrics},
  author={K{\"o}hler, Johannes and Rauch, Florian and M{\"u}ller, Matthias A and Allg{\"o}wer, Frank},
  journal={Automatica},
  volume={155},
  pages={111169},
  year={2023},
  publisher={Elsevier}
}

@inproceedings{gao2024learning,
  title={Learning-based rigid tube {MPC}},
  author={Gao, Yuning and Berberich, Julian and Allg{\"o}wer, Frank},
  booktitle={Learning for Dynamics and Control Conference},
  year={2024},
  organization={PMLR}
}

@inproceedings{fan2020deep,
  title={Bayesian learning-based adaptive control for safety critical systems},
  author={Fan, David D and Nguyen, Jennifer and Thakker, Rohan and Alatur, Nikhilesh and Agha-mohammadi, Ali-akbar and Theodorou, Evangelos A},
  booktitle={IEEE International Conference on Robotics and Automation},
  pages={4093--4099},
  year={2020},
  organization={IEEE}
}

@article{zhang2025mpc,
  title={Model predictive control with residual learning and real-time disturbance rejection: Design and experimentation},
  author={Zhang, Haoran and Liu, Yang and Wang, Jun},
  journal={Control Engineering Practice},
  volume={154},
  pages={106123},
  year={2025},
  publisher={Elsevier}
}

@article{liu2025dual,
  author    = {Liu, Yutao and Wang, Panpan and Tóth, Roland},
  title     = {Learning for Predictive Control: A Dual Gaussian Process Approach},
  journal   = {Automatica},
  volume    = {177},
  pages     = {112316},
  year      = {2025},
  publisher = {Elsevier}
}

@article{polcz2023efficient,
  author    = {Polcz, Péter and Péni, Tamás and Tóth, Roland},
  title     = {Efficient Implementation of Gaussian Process-Based Predictive Control by Quadratic Programming},
  journal   = {IET Control Theory \& Applications},
  volume    = {17},
  number    = {8},
  pages     = {943--1087},
  year      = {2023},
  publisher = {IET}
}

@article{yin2020reachability,
  author    = {Yin, He and Packard, Andrew and Arcak, Murat and Seiler, Peter},
  title     = {Reachability Analysis Using Dissipation Inequalities for Uncertain Nonlinear Systems},
  journal   = {Systems \& Control Letters},
  volume    = {120},
  pages     = {116--123},
  year      = {2020},
  publisher = {Elsevier},
  note      = {2021 Brockett-Willems Outstanding Paper Award}
}

@article{schwenkel2022model,
  title={Model predictive control for linear uncertain systems using integral quadratic constraints},
  author={Schwenkel, Lukas and K{\"o}hler, Johannes and M{\"u}ller, Matthias A and Allg{\"o}wer, Frank},
  journal={IEEE Transactions on Automatic Control},
  volume={68},
  number={1},
  pages={355--368},
  year={2022}
}

@article{liu2021random,
  title={Random features for kernel approximation: A survey on algorithms, theory, and beyond},
  author={Liu, Fanghui and Huang, Xiaolin and Chen, Yudong and Suykens, Johan AK},
  journal={IEEE Transactions on Pattern Analysis and Machine Intelligence},
  volume={44},
  number={10},
  pages={7128--7148},
  year={2021},
  publisher={IEEE}
}

@article{rahimi2007random,
  title={Random features for large-scale kernel machines},
  author={Rahimi, Ali and Recht, Benjamin},
  journal={Advances in neural information processing systems},
  volume={20},
  year={2007}
}

@article{andersson2019casadi,
  title={CasADi: a software framework for nonlinear optimization and optimal control},
  author={Andersson, Joel AE and Gillis, Joris and Horn, Greg and Rawlings, James B and Diehl, Moritz},
  journal={Mathematical Programming Computation},
  volume={11},
  number={1},
  pages={1--36},
  year={2019},
  publisher={Springer}
}

@article{wachter2006implementation,
  title={On the implementation of an interior-point filter line-search algorithm for large-scale nonlinear programming},
  author={W{\"a}chter, Andreas and Biegler, Lorenz T},
  journal={Mathematical programming},
  volume={106},
  number={1},
  pages={25--57},
  year={2006},
  publisher={Springer}
}

@article{bokor2025robust,
  author={Bokor, {\'{A}}kos M. and Biertümpfel, Felix and Seiler, Peter and Tóth, Roland},
  journal={IEEE Control Systems Letters}, 
  title={Robust Model Predictive Control for Spacecraft Rendezvous under Sector-Bounded Nonlinearities},
  year={2025},
  volume={},
  number={},
  pages={},
  doi={10.1109/LCSYS.2025.3632763},
  note={Accepted for publication in L-CSS}
}

\end{document}